\documentclass[%
superscriptaddress,
showpacs,preprintnumbers,
nofootinbib,
 amsmath,amssymb,
 aps,
 twocolumn,
prl,
nofootinbib
]{revtex4}
\usepackage{braket}
\usepackage{bm}
\usepackage{dcolumn}
\usepackage{cancel}
\usepackage[dvipdfmx,final]{graphicx}
\input{colordvi.tex}
\usepackage{longtable}
\usepackage[compat=1.1.0]{tikz-feynman}
\tikzfeynmanset{
every edge={very thick},
}

\def\bea{\begin{eqnarray}}
\def\eea{\end{eqnarray}}

\def\Mpl{M_{\rm pl}}

\newcommand\unit[1]{\,{\rm #1}}
\newcommand\eV{\unit{eV}}
\newcommand\keV{\unit{keV}}

\newcommand\GeV{\unit{GeV}}
\newcommand\TeV{\unit{TeV}}

\def\kpc{\unit{kpc}}

\usepackage{color}
\usepackage{ulem}

\begin{document}
\preprint{CTPU-PTC-18-17}
\title{Decaying axinolike dark matter: Discriminative solution to small-scale issues}
\author{Kyu Jung Bae}
\email{kyujungbae@ibs.re.kr}
\affiliation{Center for Theoretical Physics of the Universe, Institute for Basic Science (IBS), Daejeon 34126, Korea}
\author{Ayuki Kamada}
\email{akamada@ibs.re.kr}
\affiliation{Center for Theoretical Physics of the Universe, Institute for Basic Science (IBS), Daejeon 34126, Korea}
\author{Hee Jung Kim}
\email{hyzer333@kaist.ac.kr}
\affiliation{Department of Physics, KAIST, Daejeon 34141, Korea}

\date{\today}

\begin{abstract}
The latest Lyman-$\alpha$ forest data severely constrain the conventional warm dark matter solution to small-scale issues in the cold dark matter paradigm.
It has been also reported that unconstrained astrophysical processes may address the issues.
In response to this situation, we revisit the decaying dark matter solution to the issues, discussing possible signatures to discriminate decaying dark matter from astrophysical processes as a solution to small-scale issues.
We consider an axinolike particle (ALPino) decaying into an axionlike particle (ALP) and gravitino with the lifetime around the age of the Universe.
The ALPino mass is sub-PeV and slightly ($\Delta m/m\sim 10^{-4}$) larger than the gravitino mass, and thus the dark matter abundance does not alter virtually after the ALPino decays.
On the other hand, the gravitino produced from the ALPino decay obtains a kick velocity of $\sim 30 \unit{km / s}$, which is sufficiently larger than a circular velocity of dwarf galaxies to impact their dark matter distributions.
The Lyman-$\alpha$ forest constraints are relieved since only a small fraction ($\sim10$\%) of dark matter experiences the decay at that time.
Decaying dark matter is thus promoted to a viable solution to small-scale issues.
The ALPino relic abundance is determined predominantly by the decay of the lightest ordinary supersymmetric particle.
The monochromatic ALP emission from the ALPino decay is converted to $\sim 50 \GeV$ photon under the Galactic magnetic field.
The morphology of the gamma-ray flux shows a distinctive feature of the model when compared 
to decaying dark matter that directly decays into photons.
Once detected, such distinctive signals discriminate the decaying dark matter solution to small-scale issues from unconstrained astrophysical processes.

\end{abstract}


\maketitle

{\it Introduction} -- 
Cold dark matter (CDM) is a standard paradigm of the large-scale (Mpc--Gpc) structure formation of the Universe, explaining a wide range of cosmological observations such as cosmic microwave background (CMB) anisotropies~\cite{Ade:2015xua} and galaxy clustering~\cite{Alam:2016hwk}.
Nevertheless, looking to the small-scale (sub-Mpc) matter distribution of the Universe, one finds a variety of discrepancies between CDM predictions and observations (small-scale issues)~\cite{Bullock:2017xww}.
Although hydrodynamic simulations have been demonstrating that astrophysical processes may address small-scale issues~\cite{Sawala:2015cdf, Dutton:2015nvy, Wetzel:2016wro}, the implementation of subgrid astrophysical processes is still uncertain and unconstrained.
Thus it is worth investigating alternative solutions to small-scale issues.

The too-big-to-fail problem is one prominent example of small-scale issues.
$N$-body simulations have seen  about $10$ most massive subclumps in Milky-Way-size halos to be more concentrated in the inner region ($0.1$-$1 \kpc$) than the observed dwarf spheroidal galaxies~\cite{BoylanKolchin:2011de, BoylanKolchin:2011dk}.
Conventional warm dark matter (WDM) with a few keV mass is shown to resolve this discrepancy by smearing the primordial density contrast and thus delaying a halo formation below a cutoff scale through free-streaming~\cite{Lovell:2011rd}.
The free-streaming effect is maximal at matter-radiation equality and thus suppressed are not only the formation of present subgalactic halos but also matter clustering at high redshifts.
The smoothed matter distribution at $z = 3 \text{--} 5$ is probed and severely constrained by 
the recent Lyman-$\alpha$ forest data~\cite{Viel:2013apy, Baur:2015jsy, Irsic:2017ixq, Yeche:2017upn}.
The WDM solution to the too-big-to-fail problem appears {\it not} viable~\cite{Schneider:2013wwa}.

It motivates one to look for another alternative as a solution to the too-big-to-fail problem. 
It is given in the framework of decaying dark matter (DDM), where a DDM particle decays into a stable dark matter (SDM) particle and a light invisible particle.
Previous studies~\cite{Peter:2010au, Peter:2010jy, Bell:2010fk, Peter:2010sz, Aoyama:2011ba, Wang:2012eka, Wang:2013rha, Aoyama:2014tga, Wang:2014ina, Cheng:2015dga} show that the small-scale issues are mitigated when the DDM lifetime is $\Gamma^{-1} \sim t_{\rm age} \simeq 13.8 \unit{Gyr}$ (age of the Universe) and a kick velocity is $V_{\rm k} = \Delta m / m \sim 20 \text{--} 40 \unit{km / s}$.
We refer readers to Ref.~\cite{Wang:2014ina} for a DDM simulation and direct comparison of predicted circular velocities of most massive subclumps with those observed.
Instead, here, we provide a qualitative explanation for how DDM alleviates the too-big-to-fail problem.

DDM makes significant impacts on small-scale structure whose circular velocity is smaller than the kick velocity.
If the kick velocity is larger than a circular velocity, SDM particles escape from that region; if not, halo structure is not impacted.
SDM particles move from the central region ($0.1$-$1 \kpc$) of most massive subclumps, where the circular velocity is a few tens km/s, to the outer region due to the kick velocity of $V_{\rm k} \sim 20 \text{--} 40 \unit{km / s}$.
The inner DM density profile gets shallower and diffuse and resultantly the too-big-to-fail problem is mitigated.
Furthermore, less massive subclumps can even evaporate since $V_{\rm k} \sim 20 \text{--} 40 \unit{km / s}$ is larger than the maximal circular velocity.
The resultant shallower DM density profile and the reduced number of small subclumps infer that late decaying dark matter can also solve the core-cusp problem~\cite{Moore:1999gc, deBlok:2009sp} and the missing satellite problem~\cite{Moore:1999nt, Kravtsov:2009gi},%
\footnote{However, the missing satellite problem seems quite vulnerable to astrophysical processes. For example, once an empirical relation between the stellar and halo masses is extrapolated to smaller-size haloes, late decaying dark matter could result in a smaller number of subclumps than observed~\cite{Kim:2017iwr}.
Further dedicated studies are warranted.}
respectively, in a similar parameter space~\cite{Wang:2014ina}.

\begin{figure}[!h]
\centering
\includegraphics[scale=0.6]{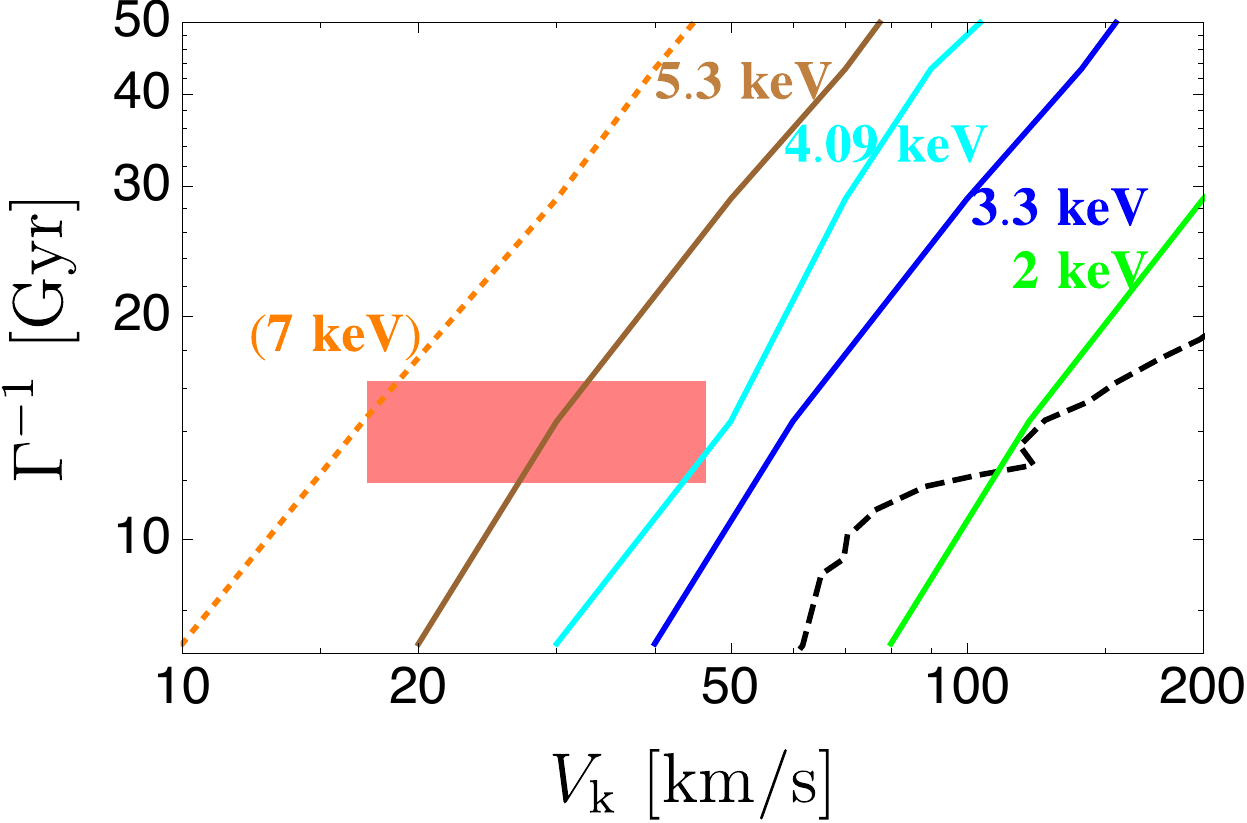}
\caption{DDM parameter space.
The red colored region indicates the parameter space suggested to be a solution to the too-big-to-fail problem~\cite{Wang:2014ina}.
The black dashed line is taken from Ref.~\cite{Wang:2013rha}, where DDM is directly examined by the Lyman-$\alpha$ forest data.
The other lines are constraints translated from the reported lower bounds on the thermal WDM mass: $m_{\rm wdm} > 2.0$ (green)~\cite{Viel:2005qj}, $3.3$ (blue)~\cite{Viel:2013apy}, $4.09$ (cyan)~\cite{Baur:2015jsy}, and $5.3 \keV$ (brown)~\cite{Irsic:2017ixq}.
We follow the procedure in Ref.~\cite{Cheng:2015dga} with the WDM transfer function in Ref.~\cite{Inoue:2014jka} and the concentration-mass relation in Ref.~\cite{Dutton:2014xda}.
One can see that once the Lyman-$\alpha$ forest constraint becomes as tight as $m_{\rm wdm} \gtrsim 7.0 \keV$, the DDM solution will be severely constrained.
}
\label{fig:late_ddm}
\end{figure}

The DDM effects on the Lyman-$\alpha$ forests are weaker than the WDM ones.
For $\Gamma^{-1} = t_{\rm age}$, only 14\% of DDM decays before $z = 3$, while 62\% of DDM decays before $z = 0$.
{\it Figure~\ref{fig:late_ddm} illustrates the viability of the DDM solution to small-scale issues in view of recent Lyman-$\alpha$ forest data}.
We have translated the reported constraints on the thermal WDM mass $m_{\rm wdm}$ into the DDM parameter space by closely following Ref.~\cite{Cheng:2015dga}.
A dedicated study like Ref.~\cite{Wang:2013rha} is preferable but beyond the scope of this paper.
We remark that this model can also easily evade constraints from the recent observations of 21~cm signal~\cite{Bowman:2018yin}
through formation of Pop-III stars and reionization~\cite{Safarzadeh:2018hhg,Schneider:2018xba}.

In contrast to the intriguing structure formation in DDM models, an investigation of the following two aspects seems missing:
\begin{itemize}
\item a concrete particle physics model interrelating $V_{\rm k} \sim 10^{-4}$ and $\Gamma^{-1} \sim t_{\rm age}$ with more fundamental parameters and providing a DDM production mechanism.
\item prospects of finding distinctive signals of a certain DDM model apart from its structure formation, making it possible to discriminate the DDM solution to small-scale issues from complex astrophysical processes~\cite{Sawala:2015cdf, Dutton:2015nvy, Wetzel:2016wro}.
\end{itemize}
Here, 
the light particle emitted from the DDM decay is the key.
It is severely constrained if it interacts electromagnetically and actually even if it is neutrino~\cite{Bell:2010fk}.
Thus one needs to introduce a feebly interacting particle beyond the standard model (SM).
On the other hand, if it does not leave any detectable signal, it does not help us to discriminate the DDM scenario as a solution to small-scale issues.
An axionlike particle (ALP) is a good candidate filling the gap.
It is naturally light and feebly interacting with SM particles.
Meanwhile, a part of the ALPs emitted from the DDM decay is converted to ${\cal O} (10) \GeV$ photons under the Galactic magnetic fields.
Depending on the ALP parameters, the resultant gamma-ray flux can be detected in existing and near future facilities such as Fermi-LAT~\cite{Ackermann:2012pya, TheFermi-LAT:2015kwa} and CTA~\cite{Consortium:2010bc}.%
\footnote{One can identify an ALP in this paper as the QCD axion, solving the strong CP problem~\cite{Peccei:1977hh, Peccei:1977ur, Weinberg:1977ma, Wilczek:1977pj}.
However, in the QCD axion case, the produced gamma-ray signal is too faint to be detected.
We will discuss this point later in the {\it ALP emission and GeV gamma ray} section.}
Remarkably its peculiar morphology over the sky makes this DDM scenario distinguishable as a solution to small-scale issues from unconstrained astrophysical processes.

If we consider an ALP as an invisible particle, it is straightforward to consider an {\it axino}like particle (ALPino) and gravitino as DDM and SDM, respectively. 
For the first compelling model, in this paper, we propose a supersymmetric (SUSY) extension of the standard model plus an ALP sector.
A fermion partner of the ALP ($a$), i.e., ALPino ($\tilde a$), is the next-to-the-lightest supersymmetric particle (NLSP) and a fermion partner of graviton, i.e., gravitino ($\psi_{\mu}$), is the LSP.
The mass of ALPino is determined by SUSY breaking and thus is expected to 
be of order the gravitino mass in supergravity
similarly to the mass of the quantum chromodynamics axino~\cite{Goto:1991gq,Chun:1992zk,Chun:1995hc}.
It is plausible that the ALPino mass is exactly the same as the gravitino mass at the tree-level, but $(m_{\tilde a} - m_{3/2}) / m_{\tilde a} \sim 10^{-4}$ ($m_{\tilde{a}}$: ALPino mass, $m_{3/2}$: gravitino mass) is achieved by one-loop correction~\cite{Moxhay:1984am, Goto:1991gq}.
{\it Intriguingly, the lifetime of ALPino is about the age of the Universe when the gravitino mass is at the sub-PeV scale and $\Delta m / m \sim 10^{-4}$.}
Such a high-scale SUSY is compatible with the $125 \GeV$ SM Higgs mass~\cite{Giudice:2011cg, Ellis:2017erg}, although it does not solve a little hierarchy or improve the grand unification since we take gaugino masses as high as PeV.
It is also encouraging that DDM is realized in SUSY models, where the parameters are well-controlled by the symmetry.
ALPino is mainly produced from the decay of the lightest ordinary supersymmetric particle (LOSP) during the reheating of the Universe.
\\

{\it ALPino Model} -- We work in a simple SUSY ALP model 
where an ALP superfield ($A$) couples to U(1) hypercharge gauge superfield $W_{B}$
and hidden SU($N$)$_{\rm h}$ gauge superfield $W_{\rm h}^{a}$.
The interactions are described by the effective superpotential given as
\bea
W_{\rm eff} &=& - \sqrt{2} g_{a B} \, A \, W_{B} W_{B} - \sqrt{2} g_{a g_{\rm h}} \, A \, W_{\rm h}^{a} W_{\rm h}^{a} \,,
\label{eq:AWW}
\eea
where $g_{aB}$ and $g_{ag_{\rm h}}$ are dimensionful coupling constants proportional to $1 / f$.
The former coupling offers production of ALPinos in the early Universe.
It also provides the ALP-photon coupling after the electroweak symmetry breaking as
\bea
{\cal L}_{\rm eff} &\supset& g_{a \gamma} \, a \, F_{\mu \nu} {\widetilde F}^{\mu \nu} \,,
\label{eq:aAA}
\eea
where $g_{a\gamma}=g_{aB}\cos^2\theta_W$ ($\theta_W$: weak mixing angle).
It converts the ALP emitted from the ALPino decay into a photon in the presence of the Galactic magnetic field.
Meanwhile, the confinement of the hidden sector at the dynamical scale $\Lambda_{\rm h}$ gives the ALP mass of $m_{a} \sim \Lambda_{\rm h}^{2} / f$ through the $g_{a g_{\rm h}}$ coupling.

The decay of ALPino into gravitino and an ALP is described by~\cite{Cremmer:1982en}
\bea
{\cal L}_{3/2} = - \frac{1}{2 \Mpl} \partial_{\nu} a \, {\bar \psi}_{\mu}  \, \gamma^{\nu} \gamma^{\mu} i \gamma_{5} \, {\tilde a} \,,
\eea
with $\Mpl = 2.43 \times 10^{18} \GeV$ being the reduced Planck mass.
It leads to~\cite{Hamaguchi:2017ihw}
\bea
\Gamma_{\tilde a}^{-1} &=& \frac{96 \pi m_{3/2}^{2} \Mpl^{2}}{m_{\tilde a}^{5}} \left( 1 - \frac{m_{3/2}}{m_{\tilde a}} \right)^{-2} \left( 1 - \frac{m_{3/2}^{2}}{m_{\tilde a}^{2}} \right)^{-3} \, \nonumber \\
&\simeq& 10 \unit{Gyr} \left( \frac{700 \TeV}{m_{\tilde a}} \right)^{3} \left( \frac{20 \unit{km/s}}{V_{\rm k}} \right)^{5} \,,
\label{eq:ALPino_decay}
\eea
with which the ALPino mass is uniquely determined by $\Gamma_{\tilde a}^{-1}$ and $V_{\rm k}$.
\\

{\it ALPino relic abundance} -- 
We can obtain the correct ALPino relic abundance, for instance, in the following low-reheating scenario.
We assume that inflaton perturbatively decays into the SM sector and its SUSY partners for simplicity.
The thermal freeze-out of the LOSP takes place during the inflaton domination before the reheating, $T_{\rm R} < T_{\rm fo} \sim m_{\rm losp}/20$, if the maximum temperature is higher than $m_{\rm losp}$.
The LOSP eventually decays predominantly into ALPino.
Other contributions such as production through scattering and decay processes of thermal particles are negligible~\cite{BKK}.
The ALPino yield is given by
\bea
Y_{\tilde a} \simeq Y_{\rm losp}^{\rm fo}\times
  \frac{4 + p}{1 + p} \, \left[ \frac{g_{*} (T_{\rm R})}{g_{*} (T_{\rm fo})} \right]^{1/2} \left( \frac{T_{\rm R}}{T_{\rm fo}} \right)^{3} \,,  
\eea
where $Y_{\rm losp}^{\rm fo}$ is the LOSP yield for $T_{\rm R} > T_{\rm fo}$ and the additional factor represents the dilution of the relic abundance during the reheating.
The LOSP yield is given by
\bea
Y_{\rm losp}^{\rm fo}\simeq \frac{(1 + p) \, H (T_{\rm fo})}{\langle \sigma v \rangle (T_{\rm fo}) \, s (T_{\rm fo})} \,,
\eea
with $g_{*} (T)$ being the effective number of massless degrees of freedom, $s (T)$ being the entropy density, and $H (T)$ being the Hubble expansion rate.
The thermally averaged annihilation cross section, $\langle \sigma v_{\rm rel} \rangle \propto v_{\rm rel}^{2 p}$, depends on a detailed SUSY mass spectrum.

For instance, we take $Y_{\rm losp}^{\rm fo} = 4 \times 10^{-13} \, (m_{\rm losp} / 1 \TeV)$ and $p = 0$ for simplicity, having in mind a fermion SUSY partner of the weak charged particles (e.g., wino and Higgsino) as the LOSP~\cite{Jungman:1995df}.
By equating the ALPino relic abundance with the observed DM mass density, one finds
\bea
T_{\rm R} \simeq 570 \GeV \left( \frac{m_{\rm losp}}{10^{6} \GeV} \right)^{2/3} \,.
\eea
Due to the low-reheating temperature, we can safely ignore the thermally produced gravitino abundance~\cite{Bolz:2000fu, Pradler:2006qh, Rychkov:2007uq}.
We find that the hidden sector quasi-stable glueball abundance produced from the SM (+ its SUSY partner) thermal bath and from the decaying sALP (scalar partner of ALP) coherent oscillation is negligible for $g_{a \gamma} \gtrsim 10^{-15}\GeV^{-1}$~\cite{BKK}.
We have also checked that the ALP abundance via the misalignment mechanism is subdominant.
Furthermore, in the parameter space of interest, the LOSP lifetime is short enough not to spoil the SM success of the big bang nucleosynthesis and not to dominate the energy density of the Universe.
\\

{\it ALP emission and GeV gamma ray} -- 
The ALPino decay produces an ALP with the energy of
\bea
E_{a} = m V_{\rm k} = 47 \GeV \left( \frac{m_{\tilde a}}{700 \TeV} \right) \left( \frac{V_{\rm k}}{20 \unit{km / s}} \right) \,,
\eea
which is also determined by $\Gamma_{\tilde a}^{-1}$ and $V_{\rm k}$.
The emitted ALP inside our galaxy is converted to the GeV gamma ray under the Galactic magnetic field.
The flux is given by
\bea
E^{2}_{\gamma} \frac{d^{2}\Phi^{a}_{\gamma}}{dE_{\gamma} d\Omega} &\simeq& 6 \times 10^{3} \, J_{\rm D, ROI} 
\, e^{- \Gamma_{\tilde{a}} t_{\rm age}} \unit{MeV / cm^{2} / s / sr} \nonumber \\
&& \times \left( \frac{E_{\gamma}}{47 \GeV} \right)^{2} \left( \frac{700 \TeV}{m_{\tilde a}} \right) \left( \frac{\Gamma_{\tilde a}}{10 \unit{Gyr}} \right) \nonumber \\
&& \times \left( \frac{1 \GeV}{\Delta E} \right) \left( \frac{1 \unit{sr}}{\Delta \Omega_{\rm ROI}} \right)  \,,
\eea
at the position of the Sun.
Here 
$E_{\gamma}$($= E_{a}$) is the energy of the converted photon
and $\Delta E$ is the energy bin size of the observation of interest.
The $J_{\rm D, ROI}$ factor is given by
\bea
J_{\rm D, ROI} = \int_{\rm ROI} d\Omega \int_{\rm los} ds \, 
P_{a\gamma}(s,\Omega)\frac{\rho (s, \Omega)}{r_{\odot} \rho_{\odot}} \,,
\label{eq:Jfactor}
\eea
with the sky region of interest (ROI) and the line of sight (los).
Here, $P_{a\gamma}(s,\Omega)$ is the local ALP-photon conversion probability,
and $\rho(s,\Omega)$ is the local dark matter density.
One can compare it with the observed diffuse gamma-ray flux in Fermi-LAT~\cite{Ackermann:2012pya}:
\bea
E^{2}_{\gamma} \frac{d^{2}\Phi^{\rm obs}_{\gamma}}{dE_{\gamma} d\Omega} \simeq 6 \times 10^{-4} \unit{MeV / cm^{2} / s / sr} \,,
\label{eq:FermiLAT}
\eea
with the ROI being $l = 0 \text{--} 360^{\circ}$ and $|b| = 8 \text{--} 90^{\circ}$ in the Galactic coordinate and the energy bin being $E = 30 \text{--} 50 \GeV$.
In this ROI,  we have $\Delta E = 20 \GeV$, $\Delta \Omega_{\rm ROI} = 10.8 \unit{sr}$, and if $P_{a\gamma}$ is constant, $J_{\rm D, ROI} \simeq 22\times P_{a\gamma}$ virtually independently of the DM profile.
It is clear that even a tiny conversion of ALP into photon, e.g., $P_{a \gamma} \simeq 4 \times 10^{-6}$, leaves an observable signal.

The ALP-photon conversion under the magnetic field is discussed 
in the literature~\cite{Raffelt:1987im, Raffelt:1996wa}.
Here we focus on the so-called adiabatic limit, where the scale lengths of the magnetic field and of electron distribution, and the propagation distance are much longer than the oscillation length~\cite{Raffelt:1996wa}.
For the propagation in the Galactic magnetic field, the formers are of order $1 \kpc$, while the latter is
\bea
0.4 \unit{kpc} \left( \frac{10^{-7} \eV}{m_{a}} \right)^{2} \left( \frac{E_{a}}{47 \GeV} \right)\,,
\label{eq:conversionprob}
\eea
for $m_{a}^{2} / E_{a} \gg g_{a\gamma} |B_{T}|$, where $B_{T}$ is the component of the magnetic field transverse to the line of sight.
It follows that we consider $m_{a} \gtrsim 5 \times 10^{-8} \eV$.%
\footnote{In this mass range, the most stringent bound for ALP-photon coupling is obtained by the number ratio of horizontal branch stars over red giants, and is given by $g_{a\gamma} \lesssim 6.6\times 10^{-11} \GeV^{-1}$~\cite{Ayala:2014pea}.
We refer readers to Refs.~\cite{Meyer:2016wrm, Vogel:2017fmc} for future prospects of covering this parameter region by ALP-photon oscillation features in gamma-ray spectra.}
The conversion probability of the propagating ALP from a given Galactic position to the Sun is given by
\bea
P_{a \gamma} (s, \Omega)
&\simeq& 2\times10^{-7}\left| \frac{B_{T} \left(s,\Omega\right) }{\mu{\rm G}} \right|^{2}\left(\frac{10^{-7}\eV}{m_{a}}\right)^{4} \nonumber \\
&&\times \left(\frac{g_{a\gamma}}{10^{-11} \GeV^{-1}}\right)^{2}\left(\frac{E_{\gamma}}{47 \GeV}\right)^{2} \,,
\label{eq:conversionprob}
\eea
which can be as large as $P_{a \gamma} \simeq 4 \times 10^{-6}$ obtained above.%
\footnote{In the QCD axion case, the conversion probability is rather small since $m_{a} \sim 10^{-2} \eV$ for $g_{a \gamma} \sim 10^{-11} \GeV$.
To obtain $m_{a} \sim 10^{-7} \eV$ for $g_{a \gamma} \sim 10^{-11} \GeV$, one needs to take the hidden dynamical scale smaller than the QCD one, $\Lambda_{\rm h} \sim 150 \keV$.}

\begin{figure}[!h]
\centering
\includegraphics[scale=0.2]{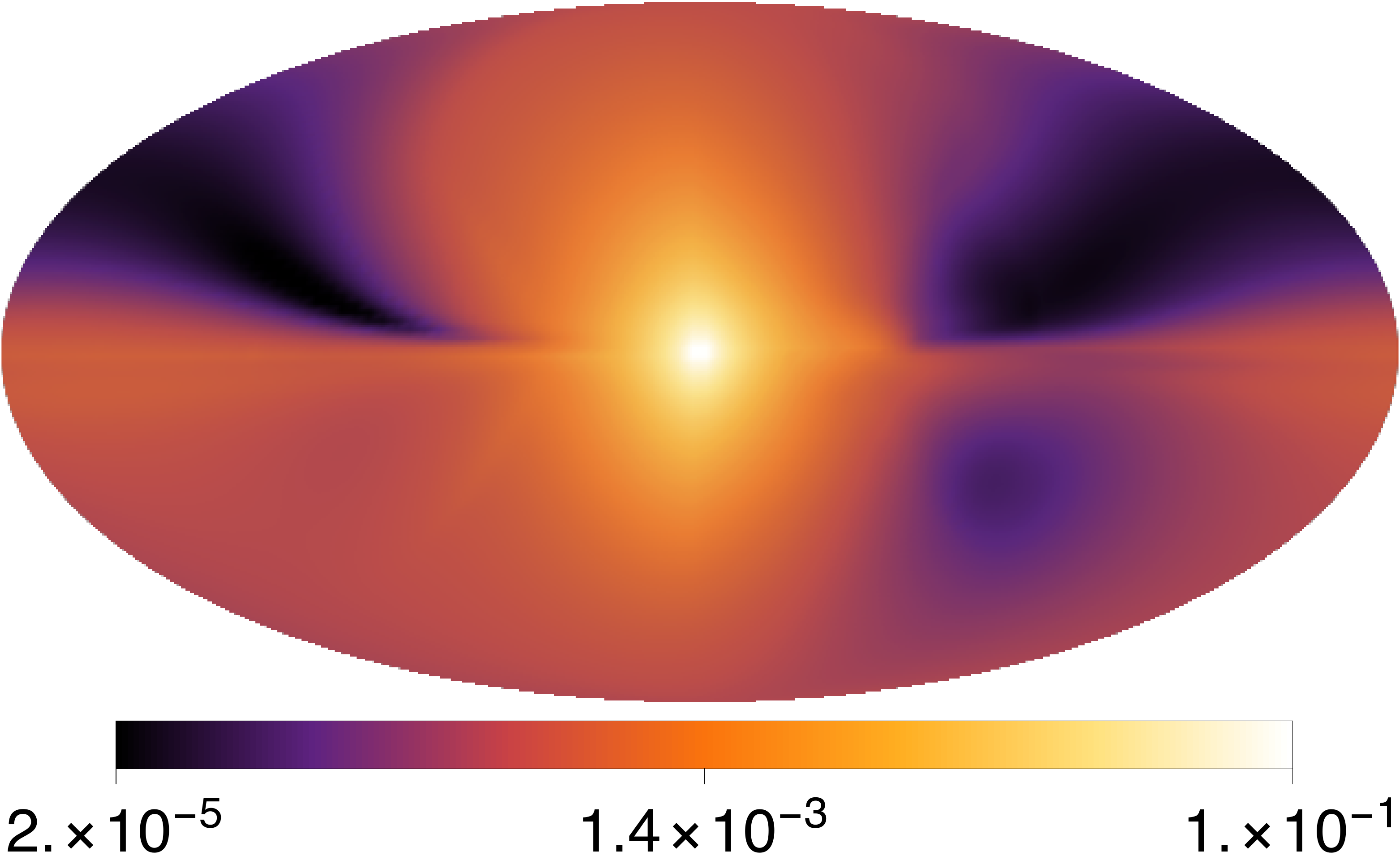}
\includegraphics[scale=0.2]{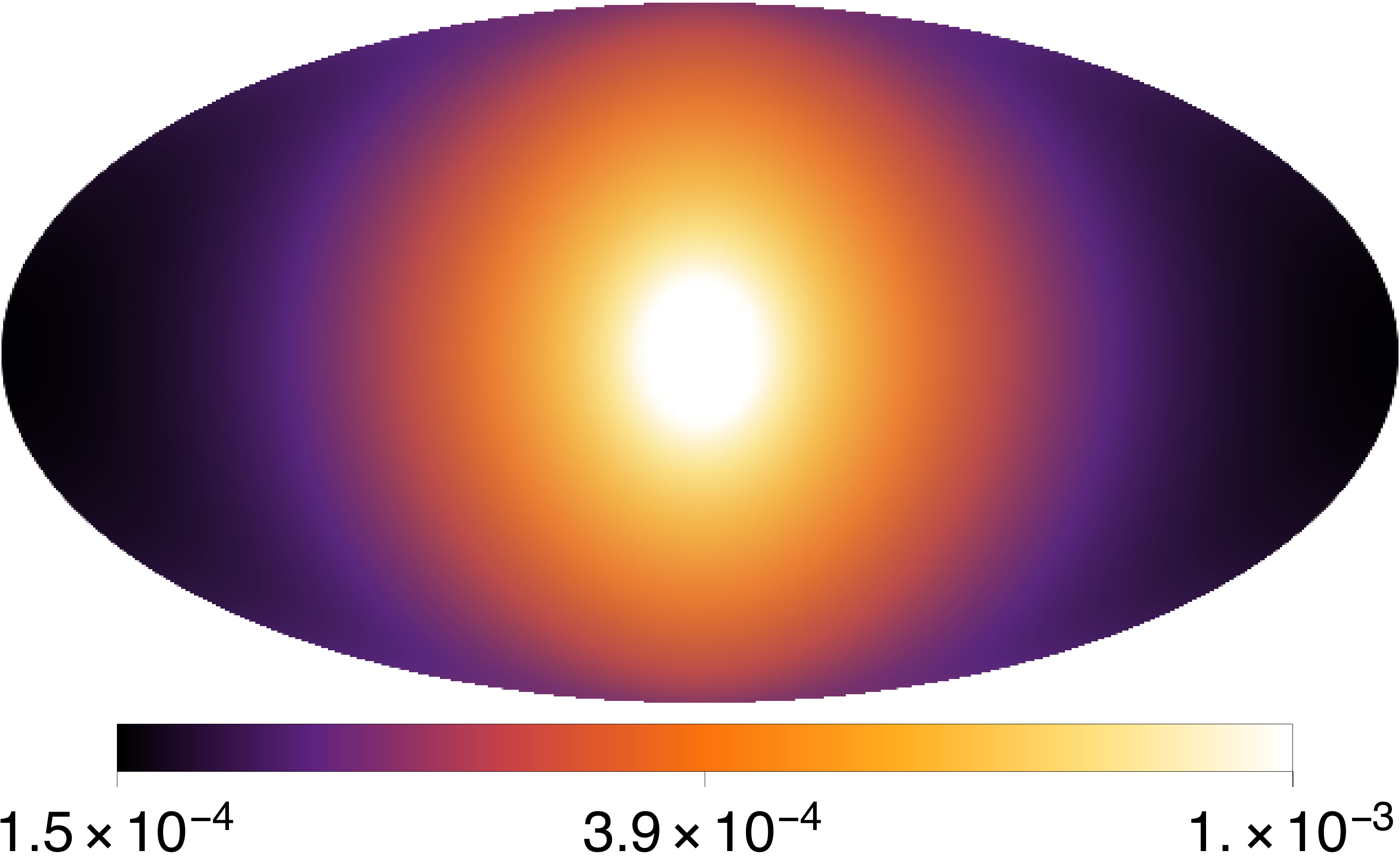}
\caption{Top: Mollweide projected sky map of gamma-ray flux converted from ALP emitted from ALPino decay.
We take $m_a=10^{-7}\,{\rm eV}$, $g_{a\gamma}=10^{-11}\GeV^{-1}$, $\Delta E_{\rm bin}=20\,{\rm GeV}$, and $E_{\gamma}=47\GeV$.
We use the Galactic magnetic field profile introduced in Ref.~\cite{Jansson:2012pc} and the NFW dark matter profile~\cite{Navarro:1995iw, Navarro:1996gj}.
The color legend denotes $E^{2}_{\gamma} d^{2}\Phi_{\gamma} / (dE_{\gamma} d\Omega)$ in units of ${\rm MeV / cm^{2} / s / sr}$.
Bottom: The same as the top panel but for DM decaying into two photons.
We set the lifetime $10^{28} \unit{s}$ and the DM mass $94 \GeV$.
}
\label{fig:morph}
\end{figure}

The $J_{\rm D, ROI}$ factor in Eq.~\eqref{eq:Jfactor} is then obtained by convoluting the DM density field and the conversion probability,
so it depends not only on the DM profile but also on the Galactic magnetic field profile.
The top panel of Fig.~\ref{fig:morph} depicts the morphology of the converted gamma-ray flux.
One can compare it with the morphology of the gamma-ray flux in a model with DM decaying into two photons 
shown in the bottom panel of Fig.~\ref{fig:morph}.
These two morphologies are clearly different since the former is a convolution of the DM density profile and the conversion probability, while the latter traces just the DM density profile.
Note that the top panel is also different from the morphology found in Refs.~\cite{Horns:2012kw, Vogel:2017fmc} since they consider an extragalactic isotropic ALP background and thus their morphology traces just the conversion probability.
We use the Galactic magnetic field model of Ref.~\cite{Jansson:2012pc} in Fig.~\ref{fig:morph} (Top panel).
We remark that morphologies in other Galactic magnetic field models are also distinctive from that for DM decaying into two photons.
Details are discussed in Appendix.

In order to obtain a robust constraint on the ALP parameter space from existing Fermi-LAT data~\cite{Ackermann:2012pya, TheFermi-LAT:2015kwa} and discuss prospects in future facilities~\cite{Consortium:2010bc}, one needs to perform a dedicated numerical analysis.
Since the adiabatic approximation is not held for some parameter space, one needs to numerically follow the ALP-photon conversion along the line-of-sights.
Uncertainty in modeling of Galactic magnetic fields also needs to be taken into account.
Such a dedicated study will be given in the future work~\cite{BKK}.
\\

{\it Concluding Remarks} -- 
We have revisited the DDM solution to small-scale issues arising in the CDM paradigm.
The DDM evades the latest Lyman-$\alpha$ forest constraints that disfavor the conventional WDM solution to the issues.
We have provided a compelling particle physics realization for the first time to our best knowledge.
We have considered an ALPino decaying into slightly lighter gravitino and an ALP with the lifetime $\Gamma_{\tilde a}^{-1} \simeq 10 \unit{Gyr}$ and the kick velocity $V_{\rm k} \simeq 30 \unit{km / s}$.
The sub-PeV ALPino mass is predicted by $\Gamma_{\tilde a}^{-1}$ and $V_{\rm k}$, while PeV SUSY breaking is compatible with the measured SM Higgs mass.
We can obtain the correct ALPino relic abundance from the LOSP decay after its freeze-out.

Not only the sub-PeV ALPino mass, but also the energy of an ALP emitted by the ALPino decay is uniquely determined by $V_{\rm k}$ as $E_{\gamma} = m V_{\rm k} \simeq 50 \GeV$.
The ALP is converted to a photon during the propagation in the Galactic magnetic field. 
We have stressed that the ALP emitted from the ALPino decay shows a unique signature that can be distinguished from usual DM decay into photon pair.
Such signatures, in principle, discriminate the DDM solution to small-scale issues from others such as unconstrained astrophysical processes.
\\

{\it Acknowledgements} -- We would like to thank Seokhoon Yun and Kohei Kamada for fruitful discussions.
This work was supported by Institute for Basic Science under the project code, IBS-R018-D1.

\appendix
\section{Appendix: Different models of Galactic magnetic fields}
The morphology of gamma-ray flux converted from ALP depends on the modeling of the Galactic magnetic field, as depicted in Eq.~\eqref{eq:conversionprob}.
In the top panel of Fig.~\ref{fig:morph}, we showed the morphology induced by the model of Ref.~\cite{Jansson:2012pc}.
However, due to the limited data sets and limited knowledge on intergalactic/interstellar medium, there are uncertainties on estimating the Galactic magnetic field, and thus the modeling of the Galactic magnetic fields is not unique.
Several models and their best-fit parameters are introduced in Refs.~\cite{Stanev:1996qj, Sun:2007mx, Pshirkov:2011um, Jansson:2012pc}.

\begin{figure}[!h]
\centering
\includegraphics[scale=0.2]{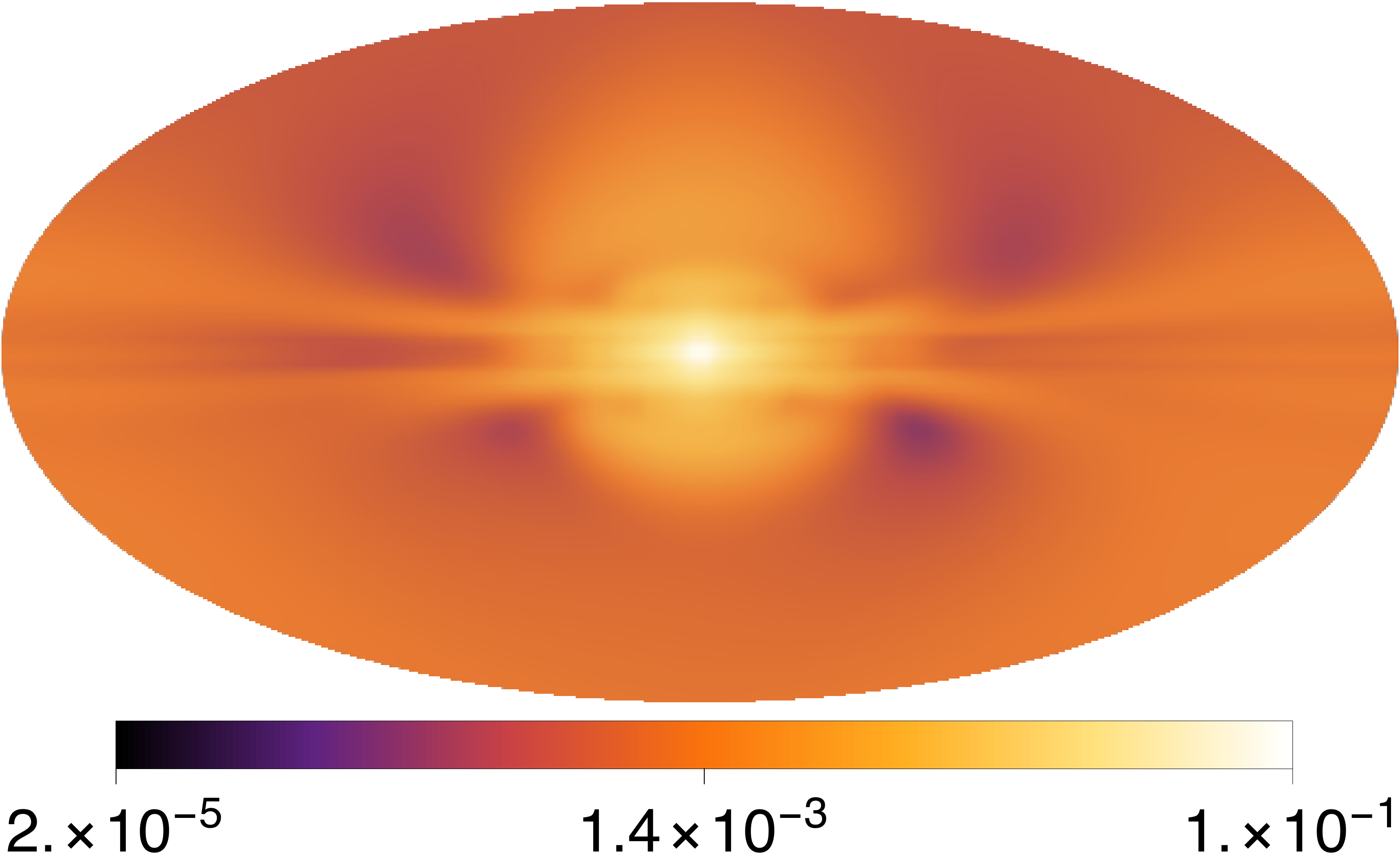}
\includegraphics[scale=0.2]{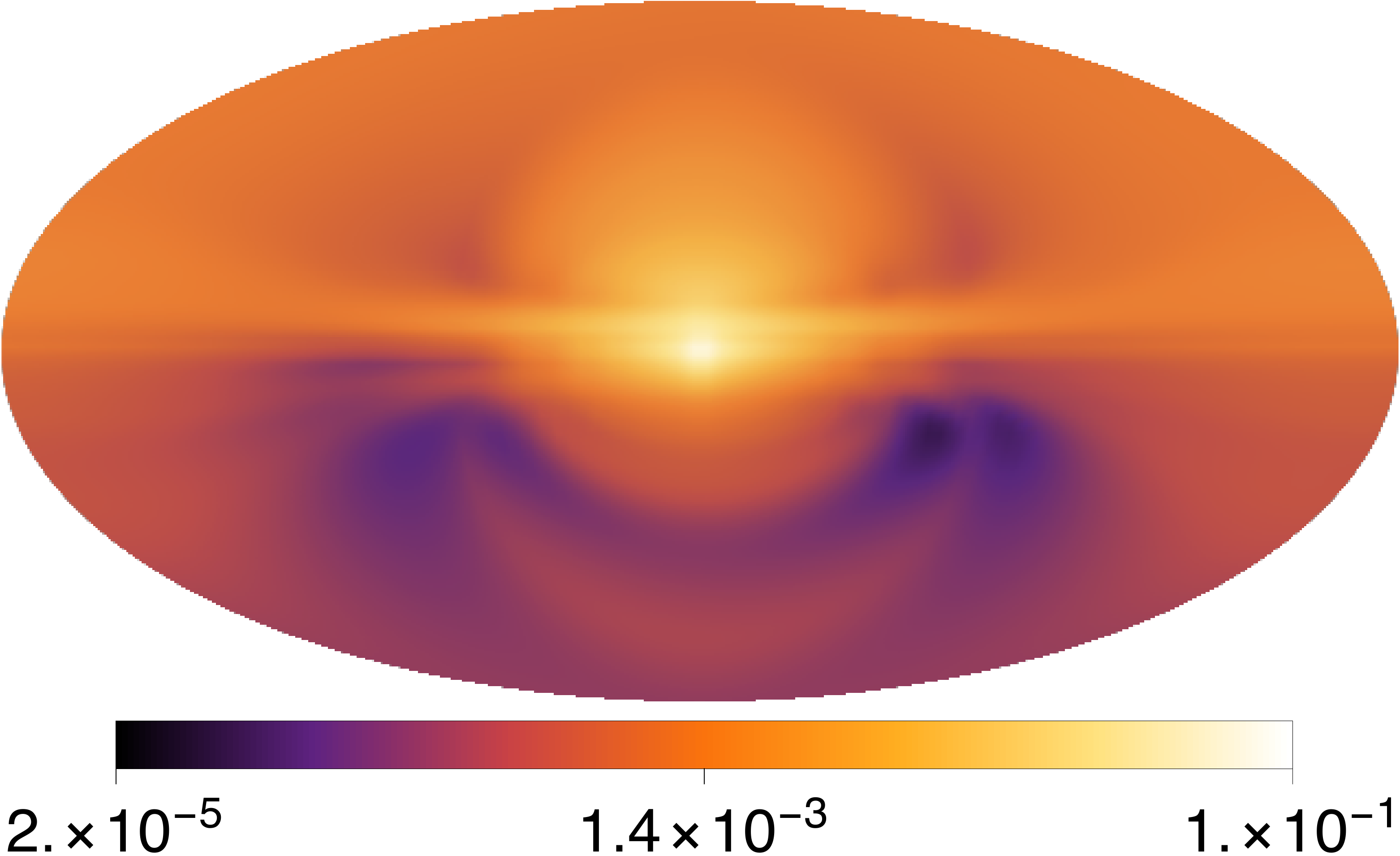}
\caption{Mollweide projected sky map of gamma-ray flux converted from ALP emitted from ALPino decay.
The parameters taken and the color legend are the same as in the top panel of Fig.~\ref{fig:morph}.
Top: The Galactic magnetic field profile introduced in Ref.~\cite{Stanev:1996qj, Pshirkov:2011um} is used.
Bottom: The Galactic magnetic field profile introduced in Ref.~\cite{Sun:2007mx, Pshirkov:2011um} is used.
}
\label{fig:morph2}
\end{figure}

In Fig.~\ref{fig:morph2}, we present the converted gamma-ray flux by taking the two models introduced in Refs.~\cite{Stanev:1996qj, Sun:2007mx, Pshirkov:2011um}.
The morphology of the converted gamma-ray flux in Fig.~\ref{fig:morph2} is still clearly different from the case of DM decaying into two photons shown in the bottom panel of Fig.~\ref{fig:morph}.
Notice that the model of Ref.~\cite{Jansson:2012pc} (Fig.~\ref{fig:morph}, top panel) exhibits a morphology with a higher contrast in intensity than those of Refs.~\cite{Stanev:1996qj, Sun:2007mx, Pshirkov:2011um} (Fig.~\ref{fig:morph2}).
This is because the model of Ref.~\cite{Jansson:2012pc} is featured by an extra ``X-shaped" component magnetic field motivated from the radio observations of external edge-on galaxies.

\bibliography{decaying_alpino}

\end{document}